# Tailoring Terahertz Propagation by Phase and Amplitude Control in Metasurfaces

Jingjing Zheng,[1,2] Xueqian Zhang,[3,*] Lixiang Liu,[4] Quan Li,[5] Leena Singh,[1] Jiaguang Han,[3] Fengping Yan,[2] and Weili Zhang[1,3]

[1]School of Electrical and Computer Engineering, Oklahoma State University, Stillwater, Oklahoma 74078, USA
[2]Institute of Lightwave Technology, Beijing Jiaotong University, Beijing 100044, China
[3]Center for Terahertz waves and College of Precision Instrument and Optoelectronics Engineering, Tianjin University and the Key Laboratory of Optoelectronics Information and Technology (Ministry of Education), Tianjin 300072, China
[4]Institute of Solid State Physics, Shanxi Datong University, Datong 037009, China
[5]School of Electronic Engineering, Tianjin University of Technology and Education, Tianjin 300222, China
[*]*Corresponding author*: alearn1988@tju.edu.cn

**Abstract**: Using metasurfaces to control the wave propagation at will has been very successful over the broad electromagnetic spectrum in recent years. By encoding specially designed abrupt changes of electromagnetic parameters into metasurfaces, such as phase and amplitude, nearly arbitrary control over the output wavefronts could be realized. Constituted by a single- or few-layer of planar structures, metasurfaces are straightforward in design and fabrication, thus promising many realistic applications. Moreover, such control concept can be further extended to the surface wave regime. In this review, we present our recent progress on metasurfaces capable of tailoring the propagation of both free-space and surface terahertz waves. Following an introduction of the basic concept and theory, a number of unique terahertz metasurfaces are presented, showing the ability in devising ultra-thin and compact functional terahertz components.

## 1 Introduction

Metamaterials, constituted by artificial metal or dielectric resonant structures with subwavelength scale, have continuously attracted enormous interest over the past decades. Their electromagnetic properties can be freely manipulated by engineering the shapes, geometric parameters and arrangement styles of the unit building blocks. Many compelling properties that are unavailable in nature can be achieved in metamaterials, such as negative refractive index [1-5], perfect lensing [6,7] and invisibility cloaking [8-12], etc. However, owning to the high loss and three-dimensional (3D) fabrication challenges [13], realistic applications of metamaterials have long been limited.

Being a 2D or quasi-2D version of metamaterials, metasurfaces intrinsically overcome the problems metamaterials have encountered. So far, many metasurface devices, including filters [14-17], polarizers [18-22], anti-reflection coating [23-25], perfect absorbers [26,27], and so on, have been successively demonstrated. Recently, metasurfaces with locally controllable ability in phase and amplitude at material interfaces have further promoted their functionality from spectral modulations into controlling the wave propagation behaviors [28-40]. The most intriguing accomplishment should be the anomalous reflection and refraction, which break the constraint of conventional Snell's laws that govern the propagation properties of electromagnetic waves at the interfaces of not only natural materials but also artificial metamaterials.

The concept of metasurfaces was proposed and discussed in detail in the optical regime [28]. With sophisticated design of spatially varied abrupt phase changes at the interface using metallic resonators, anomalous reflection and refraction that governed by the generalized Snell's laws were demonstrated. Based on such strategy, control of the output wavefront no longer necessarily requires accumulated phase through designing the refractive index and interface shape of the bulk media. Various functional devices with the phase control strategy have been investigated, such as Huygens' surface [41-43], ultra-thin flat lensing [44-51], polarizers [52,53], holograms [49,54-57], surface plasmon couplers [58-62], invisibility cloaks [63,64], achromatic devices [65-68], and many other useful components [69-75]. Moreover, the phase control concept has also given rise to other controlling concepts, such as amplitude control, polarization control and nonlinear phase control, which allow to fully manipulate the wave propagation and realize enhanced functionalities, such as meta-gratings [76,77], high-quality holography [78-80], nonlinear controls [81-85], vector beam engineering [86-88], and polarization dependent components [89-96].

Metasurfaces are becoming a general approach that can be applied nearly in the entire spectral range of electromagnetic waves to develop integrated devices, including the spectroscopically important terahertz regime, where functional devices are still extremely lacking. In this article, we present a review of our recent progress in metasurface-mediated propagation control of free-space and surface terahertz waves [31,76,97]. First, we introduce the concept and deviation of the generalized Snell's laws with basic phase control. Second, we show the realization of entire abrupt phase control using a C-shape metamaterial resonator. Then, a broadband terahertz deflector is demonstrated based on such C-shape building blocks that can bend the terahertz waves of different frequencies into various directions [31]. Next, we present broadband terahertz meta-gratings capable of controlling the number and order of the diffraction beams by incorporating abrupt amplitude change into the metasurfaces [76]. At last, we move onto the terahertz surface wave regime, and introduce an approach to control the launching phase of the terahertz surface wave by using a slit-pair resonator, which is further applied in controlling the wavefront of the surface wave [97]. The presented structures and controlling strategies are useful in realizing compact and planner terahertz spatial wave modulators and on-chip devices.

## 2 Generalized laws for reflection and refraction

When an electromagnetic wave is incident onto an interface from material $n_i$ to material $n_t$ ($n_i$ and $n_t$ are the corresponding refractive index), reflection and refraction that obey the well-known Snell's laws will take place, where $n_i\sin\theta_i = n_i\sin\theta_r$ for reflection and $n_i\sin\theta_i = n_t\sin\theta_t$ for refraction with $\theta_i$, $\theta_r$ and $\theta_t$ being the incident angle, reflection angle and refraction angle, respectively. However, if one incorporates a well-designed abrupt phase profile with uniform amplitude into the interface, anomalous reflection and refraction phenomena which no longer obey the Snell's laws will happen. Here, the abrupt phase profile can cause a sudden phase change to the incident wave when it escapes from the interface, which is distinct from the accumulation phase that determined by the wave propagation distance. In this case, additional amendment should be made to the Snell's laws.

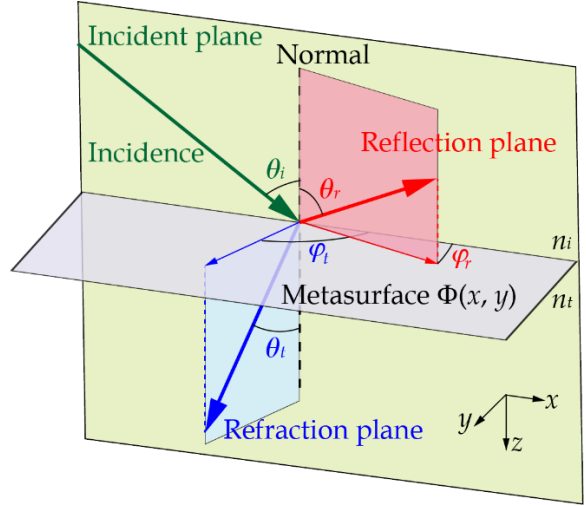

**Fig. 1.** Schematic view of the anomalous reflection and refraction effect at an interface with abrupt phase profile $\Phi(x, y)$.

Figure 1 illustrates a schematic view of anomalous reflection and refraction at an interface with $\Phi(x, y)$ abrupt phase profile. The wave vectors of the three waves can be expressed by:

$$\begin{aligned} k_{jx} &= k_0 n_j \sin\theta_j \cos\varphi_j, \\ k_{jy} &= k_0 n_j \sin\theta_j \sin\varphi_j, \\ k_{jz} &= k_0 n_j \cos\theta_j, \end{aligned} \quad (1)$$

where $j = \{i, r, t\}$ represents the incident, reflection and refraction, respectively; $k_0$ represents the vacuum wave number; $\varphi_j$ represents the azimuthal angle. Here, the incident plane is supposed in the $xoz$ plane, so $\varphi_i = 0°$. Regardless of the form of the interface, the electric and magnetic fields at the two sides of the interface should always fulfill the boundary conditions. Therefore, the in-plane components of the input and output wave vectors should always be continuous, namely, $\vec{k}_{il} + \vec{k}_{\Phi l} = \vec{k}_{jl}$ with $j = \{r, t\}$, $l = \{x, y\}$ and $\vec{k}_{\Phi} = (d\Phi/dx, d\Phi/dy)$. Here, $\vec{k}_{\Phi}$ is the additional in-plane wave vector that induced by the abrupt phase profile $\Phi(x, y)$. In this case, by substituting Eq. (1) into the continuous condition, we have

$$\begin{cases} n_i \sin\theta_r \cos\varphi_r - n_i \sin\theta_i = \dfrac{c}{2\pi f}\dfrac{d\Phi}{dx}, \\ n_i \sin\theta_r \sin\varphi_r = \dfrac{c}{2\pi f}\dfrac{d\Phi}{dy}, \end{cases} \quad (2)$$

for reflection, and

$$\begin{cases} n_t \sin\theta_t \cos\varphi_t - n_i \sin\theta_i = \dfrac{c}{2\pi f}\dfrac{d\Phi}{dx}, \\ n_t \sin\theta_t \sin\varphi_t = \dfrac{c}{2\pi f}\dfrac{d\Phi}{dy}, \end{cases} \quad (3)$$

for refraction. Equations (2) and (3) fully describe the reflection and refraction behaviors of the wave incident at such a phase-engineered interface, thus can be called as the generalized Snell's laws. It can be deduced that the value of $d\Phi/dy$ determines whether the incident, reflection and refraction planes are in the same plane. Actually, Equations (2) and (3) can also be obtained using the Fermat's principle [38]. These generalized laws indicate that the direction of wave propagation after the interface is determined by not only the refractive indices of the media but also the phase gradient of the abrupt phase profile. In particular, if further abrupt amplitude profile is incorporated into the interface, there may exist more than one phase gradient according to the Fourier decomposition. In this case, the generalized Snell's laws should be applied individually [76].

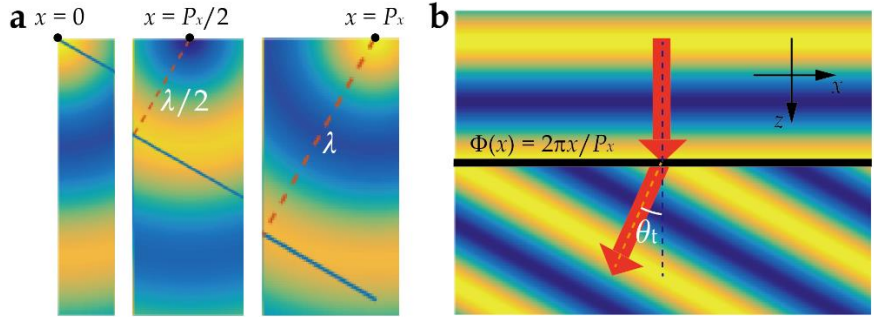

**Fig. 2. a)** Radiation field patterns of three point sources at $x = 0$, $P_x/2$ and $P_x$ with an initial phase of 0, $\pi$ and $2\pi$, respectively. **b)** Schematic of the field distribution of an anomalous refraction effect at an interface with linear abrupt phase profile $\Phi(x)$ under normal incidence.

The function of the abrupt phase profile could be understood from the view of the Huygens-Fresnel principle [98]. Here, we take a linear abrupt phase profile along the $x$ direction $\Phi(x) = 2\pi x/P_x$ as an example, where $P_x$ represents the physical distance of $2\pi$ phase change. When a normally incident plane wave ($\theta_i = 0°$) arrives at the interface with $\Phi(x)$, all the points at the interface can be regarded as new point sources with initial phase profile Φ(x). Figure 2a schematically illustrates the radiation field patterns of three point sources at $x = 0$, $P_x/2$ and $P_x$, respectively. Their initial phase values are 0°, 180° and 360°, respectively. It is seen that their equiphase surface is not parallel to the interface, as indicated by the inset blue lines. In this case, the superposition of the radiation fields from all the point sources will result in a tilt wavefront, which is anomalous compared to the conventional reflection or refraction effect. The tilt angle with respect to the interface is corresponding to the reflection/refraction angle. Figure 2b describes an anomalous refraction effect at such an interface, where an anomalous refraction phenomenon is clearly seen. According to Eq. (3), the refraction angle can be calculated as

$$\theta_t = \sin^{-1}\left(\frac{c}{fP_x}\right) \tag{4}$$

## 3 Metasurface design for complete abrupt phase control

In reality, it is quite difficult to realize such a phase profile continuously. However, it is possible to realize it in a discrete way that nearly does not disturb the performance. This can be done by using subwavelength scatters which could abruptly change the phase of the incident wave in a controllable manner, such as metamaterial resonators. Let us consider a resonator which only supports one resonance as an example, such as a bar resonator shown in Fig. 3a. The bar resonator forms an orientation angle $\beta$ with respect to the $x$ axis. The transmission property of a metasurface constituted by such resonators under the $x$-polarization incidence can be calculated by [99]

$$t_{xx} = 1 + \frac{\gamma_s \cos^2\beta}{if_0 - if - \gamma_s - \gamma_d}, \tag{5}$$

$$t_{yx} = \frac{\gamma_s \sin 2\beta}{2(if_0 - if - \gamma_s - \gamma_d)}, \tag{6}$$

where $t_{xx}$ and $t_{yx}$ represent the $x$- and $y$-polarization transmissions under the $x$-polarization incidence, respectively; $f_0$, $\gamma_s$ and $\gamma_d$ represent the resonance frequency, scattering loss rate and dissipation loss rate of the resonator, respectively; $f$ represents the frequency. Figure 3b plots the smith curves of the $t_{xx}$ and $t_{yx}$ as $f$ varies from 0 to ∞ when $\beta = $ -45° and 45°. It is seen that, for $t_{xx}$, the overlapping circles at $\beta = \pm 45°$ can only cover the first and fourth quadrants, while for $t_{yx}$, the circles at $\beta = \pm 45°$ can cover the entire four quadrants. Such phenomenon indicates that the entire $2\pi$ abrupt phase range of such metasurface could not be covered by the co-polarization output, but by the cross-polarization output. Therefore, it is possible to realize arbitrary abrupt phase profile for cross-polarization output. Similar results can also be obtained for the reflection property. Though this conclusion about the abrupt phase coverage is obtained using a structure with only one resonance, the result is also true for structures with two or more resonances, such as the C-shape or V-shape resonators which support symmetric and anti-symmetric resonances simultaneously [31,100]. In fact, multi-resonances will broaden the effective working frequency range. Figure 3c illustrates a schematic view of a C-shape resonator. The orientation angle (symmetric axis of the C-shape resonator) $\beta$ is set to be 45°, so as to maximize the amplitude of the cross-polarization output according to Eq. (6). By changing the radius $r$ and opening angle $\alpha$ of the C-shape resonator, one can pick out several structures that cover $\pi$ abrupt phase change for the cross-polarization output with equal amplitude. The other $\pi$ abrupt phase change could thus be obtained by flipping the sign of the orientation angle $\beta$ according to Eq. (6). With these resonators, nearly arbitrary abrupt phase profile is achieved.

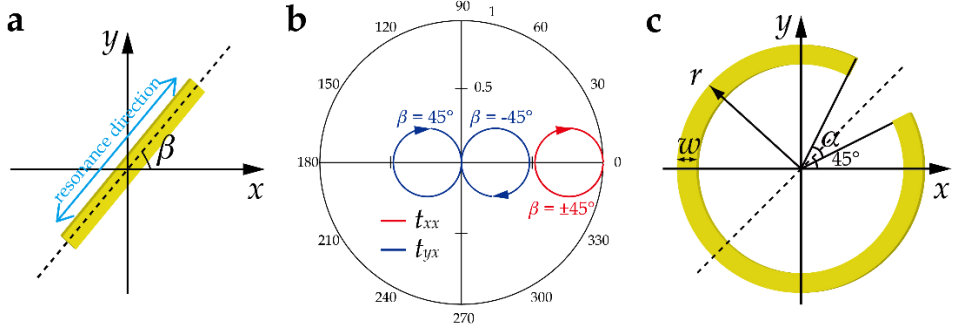

**Fig. 3. a)** Schematic of a bar resonator with only one resonance to realize the abrupt phase coverage of the output wave. **b)** Smith curve of the calculated $t_{xx}$ [Eq. (5)] and $t_{yx}$ [Eq. (6)] with $f_0 = 1.5$ THz, $\gamma_s = 0.05$ THz and $\gamma_d = 0.002$ THz. **c)** Schematic of the proposed C-shape resonator design to realize broadband terahertz abrupt phase and amplitude control.

## 4 Broadband terahertz deflector

The broadband terahertz deflector was realized using metasurfaces with linear abrupt phase gradient $d\Phi(x)/dx = 2\pi/P_x$ [31]. Eight C-shape resonators (Fig. 3c) on silicon substrate were designed based on numerical simulations with abrupt phase changes covering a $2\pi$ range with a $\pi/4$ interval. Consequently, these resonators were arranged along the $x$ axis to constitute a metasurface with desired abrupt phase gradient. A super cell of the designed metasurface and its corresponding normalized abrupt amplitude and phase distributions are illustrated in Fig. 4a. Figure 4b shows the simulated $y$-polarization transmitted field distributions of the metasurface under the normal $x$-polarization incidence at three representative frequencies 0.63, 0.8 and 1 THz, respectively. It is seen that the wavefronts of the waves are well deflected, illustrating the broadband property of the metasurface. It is also seen that the deflection angle gradually decreases with increasing frequency, which is consistent well with Eq. (4). The inset of Fig. 4c illustrates the microscope image of the fabricated metasurface. The corresponding measured results under normal incidence, +14° incidence and -15° incidence are shown in Figs. 4c to 4e, respectively, in which broadband terahertz deflection behavior is clearly observed. The solid pink lines are the calculate result using Eq. (3). It can be seen that the measured results are agreeing well with theoretical predictions, indicating a good performance of the designed terahertz deflector. Notice that, same results could also be achieved for the *x*-polarization deflection under the normal *y*-polarization incidence, owning to the structural symmetry of the C-shape resonators.

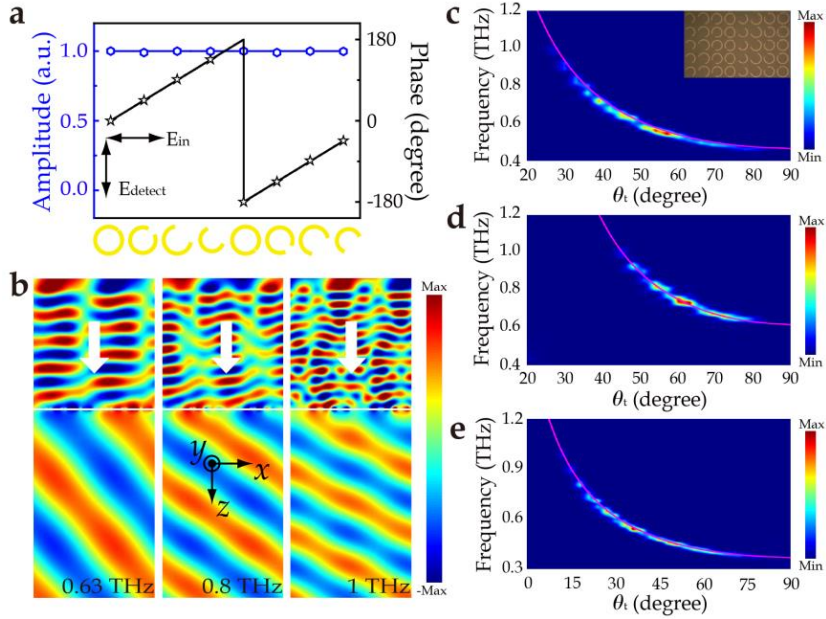

**Fig. 4. a)** Normalized abrupt amplitude and phase distributions in one periodicity of the deflector. Bottom: corresponding C-shape resonators designed to realize the desired distributions. **b)** Simulated *y*-polarization transmitted field distributions of the metasurface under the normal *x*-polarization incidence at 0.63, 0.8 and 1 THz, respectively. **c)-e)** Measured *y*-polarization output intensity of the metasurface at different deflection angles under normal, +14° and -15° incidences of the *x*-polarized wave, respectively. Inset in **c**: microscope image of the fabricated metasurface.

## 5 Broadband terahertz meta-gratings

According to Eq. (6), the amplitude of the cross-polarization output *A* can be controlled by changing the orientation angle $\beta$ of the resonator, where $A \propto |\sin(2\beta)|$. Meanwhile, the change in the orientation angle $\beta$ would not affect the phase of the output cross-polarized wave, except that a $\pi$ phase change will be incurred when the sign of $\sin(2\beta)$ changes as mentioned above. In this case, varying $\beta$ in the range of -45° to 45° is sufficient, where the abrupt amplitude achieves its maximum at $\beta = \pm 45°$ and minimum at $\beta = 0°$. By incorporating this abrupt amplitude control mechanism into the above introduced abrupt phase change control, nearly arbitrary complex transmission/reflection coefficient distribution can be achieved at the interface. This feature offers great freedom and flexibility in designing the wave propagation using metasurfaces. By using such control strategy, broadband terahertz meta-gratings that can arbitrarily control the number and order of the diffraction beams were designed [76]. The transmission of such a meta-grating should fulfill the following expression

$$t(x) = A(x)\exp[i\varphi(x)] = \sum_m A_m \exp(-i2m\pi x/d), \tag{7}$$

where $d$ is the grating periodicity, and $m$ is an integer denoting the diffraction order, $A_m$ is the amplitude of the $m$th diffraction order. For the case of a single diffraction order, the abrupt amplitude distribution is uniform and the meta-grating actually becomes a deflector as described above. For the case of two or more diffraction orders, the abrupt amplitude control becomes important. Figures 5a and 5c illustrate two meta-grating designs with two (1st and 3rd) and three (1st, 2nd, 3rd) diffraction orders of equal amplitude, respectively. The corresponding normalized amplitude and phase distributions (solid lines) in one grating periods were calculated using Eq. (7). To realize the meta-gratings, sixteen C-shape resonators were employed to constitute one supercell. Their abrupt phase responses were achieved by designing their geometric parameters, while their amplitude responses were achieved by rotating their orientation angle. Figures 5b and 5d illustrate the corresponding experimental results of the two meta-gratings under normal incidence, where two and three diffraction orders with broadband properties are clearly seen. The solid pink lines are the corresponding calculation results using Eq. (3), in which the abrupt phase gradients $d\Phi/dx$ are $\{2\pi/d, 6\pi/d\}$ and $\{2\pi/d, 4\pi/d, 6\pi/d\}$, respectively. It can be seen that the experimental results are agreeing well with the theoretical design.

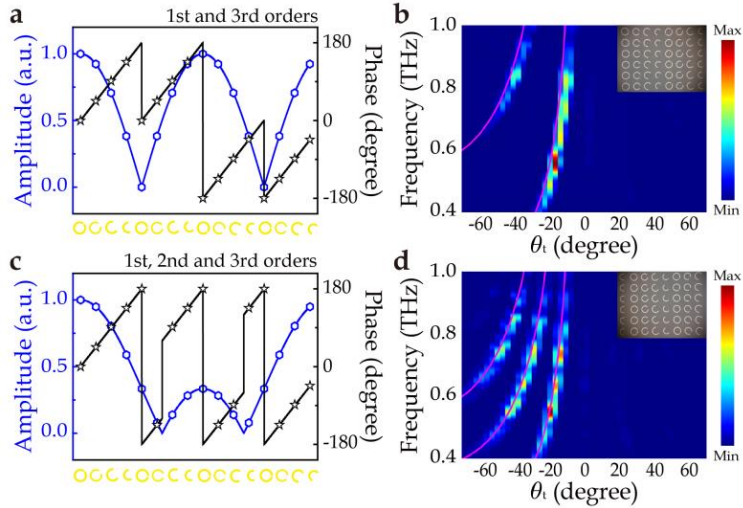

**Fig. 5. a), c)** Normalized abrupt amplitude and phase distributions in one periodicity of the meta-gratings with 1st, 3rd and 1st, 2nd, 3rd diffraction orders, respectively. Bottoms: corresponding C-shape resonators used to realize the desired distributions. **b), d)** Corresponding measured *y*-polarization output intensity at different diffraction angles of the two meta-gratings under normal incidence of the *x*-polarized waves. Insets: corresponding microscope images of the fabricated metasurfaces.

# 6 Handedness phase control for surface wave launching

Control of terahertz propagation using metasurfaces has been successfully demonstrated in the free-space regime. Here, we show that such concept could also be applied in surface wave launching and control [97]. Figure 6a illustrates a unit cell of the surface wave launcher, which contains two identical slit resonators. For a single slit resonator, it can only be excited by a polarization component perpendicular to the slit, and it radiates surface wave as an in-plane dipole. The two slits shown in Fig. 6a are perpendicular to each other with $\theta_2 - \theta_1 = 1.5\pi$ and separated with a distance $s = \lambda_{SW}/2$, where $\theta_1$ and $\theta_2$ are the angles that formed by the normal directions ($\vec{n}_1$ and $\vec{n}_2$) of the two slits and the $x$ axis; $\lambda_{SW}$ is the interested surface wave wavelength. In this case, under the circular polarization incidence, the surface wave field at point M can be expressed as:

$$\vec{E}_M = i\sqrt{2}/2 \, sign(l) A \exp(ik_{SW}|l|) e^{i2\sigma\theta_1} \hat{a}, \tag{1}$$

where $l$ is the distance between M and the center of the two slits; $A$ is the coupling coefficient from the incident wave to surface wave; $k_{SW}$ is the surface wave wavenumber; $\sigma$ is the handedness of the circular polarization with $\sigma = -1$ and $+1$ representing the left-handed circular polarization (LCP) and right-handed circular polarization (RCP), respectively; $\hat{a}$ is a unit vector that represents the direction of the surface wave field. It can be seen that the amplitude of the surface wave field is uniform at any point satisfying $|l| \geq s/2$, whereas the initial phase is widely adjustable by changing the orientation angle of the slits, and the sign of the phase is determined by the handedness of the incident circular polarization. With such slit-pair resonator, similar propagation control that presented above could also be realized in the surface wave regime. By arranging the slit-pair resonators into a column with linear phase profile, handedness controlled anomalous surface wave launching is observed under normal incidences of the circular-polarized waves, where the directions of the launched surface waves are not perpendicular to the column but form an anomalous angle, as shown in Fig. 6b. It is seen that, under the LCP incidence, the surface wave propagates obliquely upwards, while under the RCP incidence, the surface wave propagates obliquely downwards. In contrast, by arranging the slit-pair resonators into a column with parabolic phase profile, focusing of the surface wave is observed under the LCP incidence, while diverging of the surface wave is achieved under the RCP incidence, as shown in Fig. 6c.

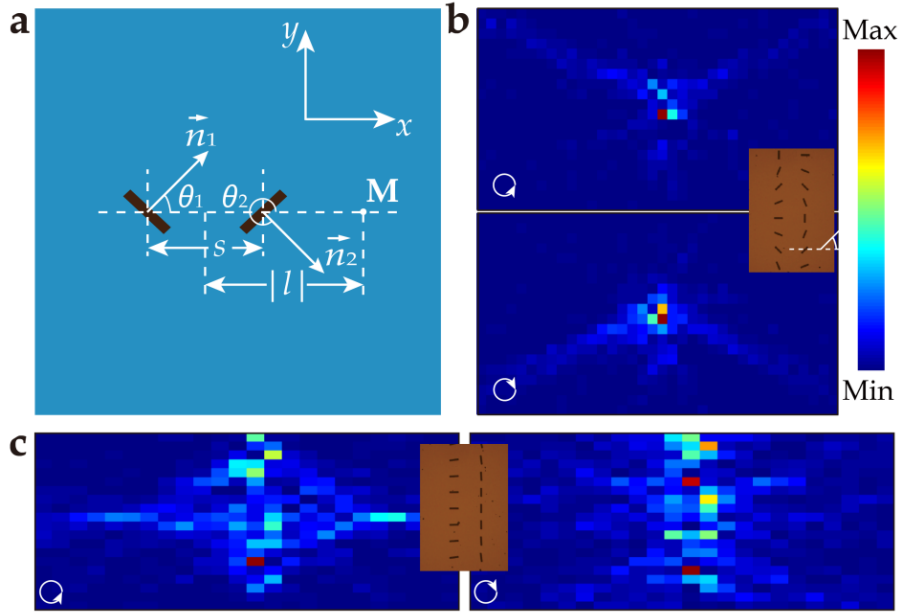

**Fig. 6. a)** Schematic view of the proposed slit-pair resonator. **b)** Measured surface wave intensity distributions of the designed column structure with linear phase profile under the LCP (upper) and RCP (lower) incidence, respectively. **c)** Measured surface wave intensity distributions of the designed column structure with parabolic phase profile under the LCP (left) and RCP (right) incidences, respectively. Insets in (b) and (c): corresponding microscope images of the fabricated structures.

## 5 Conclusion

Manipulation of free-space and surface terahertz waves using phase and amplitude control in metasurfaces were presented. The controlling strategies have recently fertilized a number of proof-of-concept functional terahertz devices, such as flat lens array for sensing the incident terahertz wavefront [47], holograms for generating high quality terahertz images [55,79], meta-gratings for realizing dual-band terahertz wave deflection [68], and active modulators for controlling the amplitude of the deflection wave [70], as well as compact devices for sensing the incident polarization state and external refractive index [97]. The demonstrated meta-device designs will be very useful in realizing ultra-thin planner terahertz wave modulators and miniature on-chip devices. Meanwhile, the design strategies may also be applicable in other regimes of the electromagnetic waves.

## Acknowledgement

The authors thank Z. Tian, J. Gu, C. Ouyang, Y. Li, Y. Xu, X. Su, N. Xu, W. Yue, M. Kenney, Y. Shi, R. Singh, and S. Zhang for their outstanding contributions and efforts in this work. The authors gratefully acknowledge the support of the National Key Basic Research Program of China (Grant No. 2014CB339800), the National Natural Science Foundation of China (Grant Nos. 61138001, 61275091 and 61420106006), the Program for Changjiang Scholars and Innovative Research Team in University (Grant No. IRT13033), the Major National Development Project of Scientific Instruments and Equipment (Grant No. 2011YQ150021), and the US National Science Foundation (Grand No. ECCS-1232081).